\documentclass[aps,prl,twocolumn,superscriptaddress,showpacs]{revtex4}

\usepackage{graphicx}
\usepackage{amsmath} 
\usepackage{amssymb}
\usepackage{array}
\usepackage{epstopdf}

\begin{document}
\def\rpdbi{$R$PdBi}
\def\tc{$T_{c}$}
\def\tn{$T_{N}$}
\def\hc2{$H_{c2}$}
\def\pdbi2{PdBi$_2$}

%\title{Non-centrosymmetric superconductivity and magnetic order in tunable topological semimetal candidate Half Heusler {\rpdbi} ($R$: rare earth) }
\title{Topological {\rpdbi} half-Heusler semimetals: \\
a new family of non-centrosymmetric magnetic superconductors}


\author{Y.~Nakajima}
\author{R.~Hu}
%\altaffiliation{Current address: Rutgers}
\author{K.~Kirshenbaum}
%\altaffiliation{Current address: Brookhaven}
\author{A.~Hughes}
\author{P.~Syers}
\author{X.~Wang}
\author{K.~Wang}
\author{R.~Wang}
\author{S.R.~Saha}
 \affiliation{Center for Nanophysics and Advanced Materials, Department of Physics, University of Maryland, College Park, MD 20742}
\author{D.~Pratt}
\author{J.W.~Lynn}
\affiliation{NIST Center for Neutron Research, National Institute of Standards and Technology, Gaithersburg, Maryland 20899}
\author{J.~Paglione}
 \email{paglione@umd.edu}
 \affiliation{Center for Nanophysics and Advanced Materials, Department of Physics, University of Maryland, College Park, MD 20742}

\date{\today}

\begin{abstract}

%The bulk-boundary correspondence that gives rise to symmetry-protected surface states in materials with non-trivial electronic topologies provides a novel platform for realizing exotic states of matter that emerge from tr

%The interplay between superconductivity and magnetism...
%provides a new opportunity to realize exotic states of matter 
%rich interplay of particle-particle and particle-hole states
%mix with new exotic states achievable through topological engineering.

We report superconductivity and magnetism in a new family of topological semimetals, the ternary half Heusler compounds {\rpdbi} ($R$ : rare earth).  
In this series, tuning of the rare earth $f$-electron component allows for simultaneous control of both lattice density via lanthanide contraction, as well as the strength of magnetic interaction via de Gennes scaling, allowing for a unique tuning of both the normal state band inversion strength, superconducting pairing and magnetically ordered ground states.
Antiferromagnetism with ordering vector ($\frac{1}{2},\frac{1}{2},\frac{1}{2}$) occurs below a N\'eel temperature that scales with de Gennes factor $dG$, while a superconducting transition is simultaneously suppressed with increasing $dG$. With superconductivity appearing in a system with non-centrosymmetric crystallographic symmetry, the possibility of spin-triplet Cooper pairing with non-trivial topology analogous to that predicted for the normal state electronic structure provides a unique and rich opportunity to realize both predicted and new exotic excitations in topological materials. 

\end{abstract}


\maketitle

%********************************
%INTRO

%Superconductivity and magnetism, traditionally antagonistic, are empirically found to share an intimate relationship in forming correlated states of matter  -add topological aspects, and provides a new avenue for investigation of exotic physics and realization of new technologies not previously considered.

Topological insulators (TIs) have recently caused a paradigm shift in the traditional classification of quantum phases of matter \cite{hasan10,ando13}. Compared to condensed matter states well understood using the concept of spontaneous symmetry breaking, the lack of symmetry-breaking in insulators with topological order, including the integer quantum Hall states, is now understood to arise from the presence of non-trivial topological components and to lead to gapless boundary modes with chirality. In the so-called Z2 2D and 3D systems, these topologically protected metallic states carry extreme interest due to their potential for realizing new technologies in spintronics and quantum computation. Combined with symmetry-breaking ordered states, TI states can give rise to unusual collective modes, such as Majorana fermions \cite{Wilcz09} with superconductivity and axions \cite{wilcz87} with magnetic order occurring in the topologically non-trivial materials. Besides the exotic collective modes, antiferromagnetism breaks time reversal and translational symmetries but preserves the combination of both symmetries, leading to a different type of TI, the antiferromagnetic TI \cite{mong10}. Despite extensive studies on bismuth-based TI materials \cite{hasan10}, only a few materials have been identified that may harbor an interplay of symmetry-breaking and topological phases \cite{erick09,hor10a,kirsh13}.

The large family of ternary half Heusler compounds \cite{graf11} is a prime candidate for combining topological and symmetry-breaking phases to explore new collective behavior.
These compounds straddle the border between topologically trivial and non-trivial electronic structure due to a strong tuning of the $s$- and $p$-like band inversion via atomic number, lattice density and spin-orbit coupling strength \cite{al-sa10,lin10,chado10}.
In addition, the stabilization of magnetic and superconducting ground states in these materials allows for a controlled interplay of symmetry and topology to be put to use.
In particular, the rare earth-based {\rpdbi} half-Heusler series is ideal for this purpose. Antiferromagnetic long range order has been observed in polycrystalline {\rpdbi} at low temperatures \cite{gofry11a}, promising emergence of antiferromagnetic topological states \cite{mong10,mulle14}. Moreover, in relation to the discovery of superconductivity in the related platinum-based compounds YPtBi \cite{butch11} and LuPtBi \cite{tafti13a}, recent reports of superconductivity in ErPdBi \cite{pan13} and LuPdBi \cite{xu14} suggest this family hosts an interesting interplay of ground states.

Here we report a systematic study of superconductivity and magnetism in single crystals of the half-Heusler series {\rpdbi} ($R$ = Y, Sm, Gd, Tb Dy, Ho, Er, Tm, and Lu) grown using a Bi self-flux technique as described in the Methods section. We find that tuning the de Gennes factor $dG$ scales local-moment N\'eel order while simultaneously suppressing the superconducting transition temperature {\tc}. 
Aspects of the superconducting state, including a possible spin-triplet pairing symmetry, confirm {\rpdbi} as a new non-centrosymmetric magnetic superconductor family.
Overall, the combination of magnetism and superconductivity together with the lack of crystallographic inversion symmetry and tunable band inversion strength in this system provides a promising route to achieving new quantum states of matter.

The MgAgAs-type crystallographic unit cell of the {\rpdbi} family is readily tuned in size by using the lanthanide contraction effect, whereby the cubic lattice parameter is continuously reduced as heavier $R$ elements are substituted. The lattice spacing has in fact been predicted as a key parameter which divides the {\rpdbi} system into trivial and non-trivial materials at the critical value of $\sim$6.62 {\AA}, where the calculated band inversion strength $\Delta E =E_{\Gamma 8} - E_{\Gamma 6}$ crosses from positive to negative ($E_{\Gamma i}$ is the energy of the bands with $\Gamma_i$ symmetry, with $\Gamma_6$ twofold and $\Gamma_8$ fourfold degenerate with total momentum $J=3/2$) \cite{chado10}. 
As shown in Fig.~1, while Sm, Y, Dy, Tb, and Gd are expected to have $\Delta E < 0$ and therefore trivial band topologies, the Ho, Er, Tm, and Lu systems with $\Delta E >0$ are expected to be non-trivial, with increasing band inversion strength as the lattice constant is further reduced. 

At the same time, substitution of different rare earth species allows fine-tuning of the magnetic properties, which are rather simple and well described by the presence (or absence) of localized $4f$ rare earth moments. Typical Curie-Weiss behavior is observed in the magnetic susceptibility, as shown in Fig.~2a, indicating the effective moments in the magnetic compounds are close to those expected for free $R^{3+}$ ion moments (see TABLE I in Supplementary Materials (SM)) with the exception of $R$=Sm, where the $J$-multiplet lying just above the ground state is very close in energy. The field dependence of the magnetization as shown in Fig.2d saturates at high field, which also indicates that the $f$-electrons are well localized. In contrast, the non-magnetic members YPdBi and LuPdBi exhibit diamagnetic behavior as shown in the inset of Fig.2a.

The $R$ local moment sublattice leads to long-range antiferromagnetic (AFM) order in $R$=Sm, Gd, Tb, and Dy, as evidenced by abrupt drops in susceptibility shown in Figs.~2b and 2c, with transition temperatures consistent with previous work on polycrystalline samples \cite{gofry11a}. Low-temperature specific heat measurements confirm the thermodynamic AFM transitions (see SM), and reveal low-temperature ordering in HoPdBi and ErPdBi at 1.9~K and 1.0~K respectively \cite{gofry11a,pan13}. We also observe a precursor of magnetic order in TmPdBi, as evidenced by a huge divergence of heat capacity reaching $\sim$ 10 J/mol~K$^{2}$ at the lowest measured temperatures.

The ordered magnetic structure for DyPdBi was determined by neutron diffraction measurements on a sample of randomly oriented crushed single crystals, and clearly reveals magnetic Bragg peaks corresponding to half-integral reflections of an fcc type-II antiferromagnet as shown in Fig.~3a. This magnetic structure is characterized by a doubling of the simple fcc Dy unit cell along all three crystallographic directions as illustrated in the inset of Fig.3b, suggesting a similar structure proposed for topological antiferromagnetism \cite{mong10,mulle14}. We note that for $R$ = Tb and Ho, we obtain the same spin structure as DyPdBi. A mean-field fit of the temperature dependence of the intensity of the $Q$ = ($\frac{1}{2},\frac{1}{2},\frac{1}{2}$) Bragg peak (Fig.~3b) results in {\tn} = 4.9 K for single crystal TbPdBi, in good agreement with the magnetic, transport and thermodynamic measurements.

Charge transport measurements at higher temperatures reveal the normal state band structure and magnetic nature of {\rpdbi}. Fig.~2e presents the semi-metallic nature of the temperature dependence of the resistivity $\rho(T)$ for {\rpdbi}, while the Hall effect carrier density $n$ at 1.8 K is very small ($n\sim 10^{19}$ cm$^{-3}$) and consistent with band calculations \cite{al-sa10,lin10,chado10}. 
%Broad peaks near 20~K - 200~K in the resistivity of magnetic rare earth {\rpdbi} are strongly suppressed by magnetic field [NOT SHOWN?], possibly associated with the suppression of spin scattering due to localized moments. 
For $R=$ Sm, Gd, Tb, Dy and Ho, an anomaly associated with magnetic order is observed in the resistivity at temperatures consistent with the thermodynamic measurements.  
Except for $R$ = Gd, all heavy $R$ members of the {\rpdbi} series undergo a superconducting transition at low temperatures as evidenced in low-temperature resistivity measurements presented in Fig.~4a. We observe a rather high transition temperature ($\sim 1.6$ K) for non-magnetic YPdBi and LuPdBi, while no trace of superconductivity is found for GdPdBi down to 20~mK. As plotted in Fig.~4c, we confirm a large diamagnetic screening in ac susceptibility measurements that onsets below {\tc}, consistent with the resistive transition temperature and with an estimated full volume fraction of the sample as determined by calibrations using aluminum. The large, but non-saturating diamagnetic screening has also been observed in other superconducting half Heusler compounds \cite{butch11,tafti13a,pan13}, attributed to the extremely long penetration depth \cite{bay14}. Note that very small traces of an impurity superconducting phase with $T_c \sim 1.5$~K but very different magnetic properties are mostly removed with proper crystal synthesis methods (see SM for details).

Interestingly, specific heat measurements for non-magnetic LuPdBi and YPdBi do not reveal a discernible signature of $T_{c}$ within the present experimental resolution, as shown in Fig.~4d. 
With strong evidence for the bulk nature of superconductivity in the half-Heuslers \cite{butch11,tafti13a,pan13,xu14}, the absence of a jump in $C(T)$ can presumably be explained by the peculiar superconductivity in this system: low carrier density, non-centrosymmetric structure, and coexistence of magnetism and superconductivity. The measured electronic component $\gamma_n$, obtained from fits to $C/T=\gamma_n +\beta T^2$, where $\beta T^2$ is a phonon contribution, is found to be $= 0.0\pm 0.5$ mJ/mol K$^{2}$ (Fig.~4d), attributed to the low carrier density and very small effective mass \cite{butch11}. An estimate of $\gamma_n$ based on the carrier density $n=10^{19}$~cm$^{-3}$ and effective mass $\sim 0.09 m_e$, where $m_e$ is electron mass \cite{wang13}, and assuming $\Delta C/\gamma_{n}T_{c} = 1.43$ for the BCS weak coupling limit, yields an expected jump at {\tc} of $<$ 0.2 mJ/mol K, beyond the resolution of our experiment (inset of Fig.~4d). 

Because non-centrosymmetry requires a mixing of singlet and triplet paring states \cite{gorko01}, the absence of a jump in $C(T_c)$ may actually be considered evidence for the presence of a dominant triplet component, in analogy with the tiny heat capacity jump measured in the A$_{1}$ phase of $^{3}$He \cite{vollh90}. 
This is corroborated by the behavior of the upper critical field \hc2, which suggests an unusual superconducting state. We plot {\hc2} for YPdBi, LuPdBi, and DyPdBi obtained from resistive transitions as a function of $T$ in Fig.~4b. \hc2 increases linearly with decreasing temperature down to $\sim T_{c}/5$, in contrast with $H_{c2}(T)$ in conventional type II superconductors determined by orbital depairing as described by the Werthamer-Helfand-Hohenberg theory, $H_{c2}(0)=-\alpha T_{c} dH_{c2}/dT|_{T=T_{c}}$, where $\alpha$ is 0.69 in the dirty limit and 0.74 in the clean limit \cite{werth66}. However, $H_{c2}(0)$ values of 2.7 T ($\alpha = 0.82$ ), 2.9 T ($\alpha = 0.91$ ), and 0.52 T ($\alpha = 0.93$) for Y, Lu, and Dy, respectively, all exceed the conventional orbital limit, which can occur for several reasons. Multiband effects and Fermi surface topology \cite{gurev03,shiba06,nakaj12} can give rise to an exceedingly large orbital depairing field, but this is not supported by the calculated \cite{al-sa10,lin10,chado10} and observed \cite{butch11,liu11} simple band structures.
%From the Mcmilan formula, we can obtain electron-boson coupling constant $\Lambda \sim$, which defy the possibility of strong coupling effect. 
As previously shown for YPtBi \cite{butch11}, the observed \hc2 curve is close in form to that of a polar $p$-wave (triplet) state \cite{schar80} as plotted in the inset of Fig.~4c. Together with the superconductivity in the clean limit $\ell > \xi$ obtained from experimental parameters, $\ell = 70$ nm and $\xi =10$ nm, the quasilinear \hc2 is suggestive of the finite triplet contribution to the pairing state due to the non-centrosymmetric nature. We note that $H_{c2}(0)$ is comparable to the Pauli paramagnetic limiting field $H_{p}$ (Tesla) = 1.84$T_{c}$ (K) obtained from a simple estimation within the weak-coupling BCS theory with $\Delta = 1.76k_{B}T_{c}$.

%%%%%%%%%%%%%%%%%%%%%%%%%%%%%%%%%%%%%%

%magnetic pair-breaking in NCS?
%[If SC is triplet in NSC, the spin-flip scattering due to magnetic impurities is in favor of the spin triplet, leading to no suppression of Tc. The potential scattering part comes from magnetic impurities, however, does suppress Tc, similar to non-magnetic scattering in nodal SC. Therefore, overall suppression of Tc in triplet NSC is sizable like that due to magnetic impurities in singlet SC.]

%%%%%%%%%%%%%%%%%%%%%%%%%%%%%%%%%%%%%%

As shown in the phase diagram (Fig.~5), the evolution of magnetism and superconductivity as a function of local moment exchange strength reveals a well-behaved trade-off and coexistence from one ground state to the other. 
We plot {\tc} and {\tn} as a function of the de Gennes factor $dG=(g_{J}-1)^{2}J(J+1)$, where $g_{J}$ is the Land\'e factor and $J$ is the total angular momentum of the $R^{3+}$ ion Hund's rule ground state. {\tn} scales well with $dG$ for RPdBi, which indicates the coupling between the conduction electrons and the local magnetic moments giving rise to the long range magnetic order is due to the RKKY interaction. More interestingly, {\tc} is suppressed almost linearly with $dG$, which indicates that magnetic rare earth $R^{3+}$ ions are a source of magnetic pair breaking \cite{mulle01}. 
%It should be noted that the reduction of {\tc} with $dG$ also suggests indirectly but strongly that the superconductivity in {\rpdbi} should be bulk in nature.

%[** ORDER PAPER THIS WAY??: ]

The superconductivity found in {\rpdbi} has several peculiar features which may give rise to novel phenomena involving topological surface states or excitations.
First, these superconductors derive from a band structure with extremely low carrier density of $\sim 10^{19}$ cm$^{-3}$, rivaled only by SrTiO$_{3-x}$ \cite{lin13a} and FeSe \cite{teras14}.
% [FeSe F~50-600T, similar or rather larger than RPdBi]
In the superconducting state, this together with a very light effective mass of 0.09$m_{e}$ \cite{wang13} leads to an extreme long penetration depth $\lambda \sim$ 1 $\mu$m; this situation can give rise to anomalous vortex states and also proximity effect to the topological surface states if present. 
More fundamentally, the low carrier density translates to a very small Fermi energy (on the order of several hundred Kelvin), putting the usual Frohlich approximation ($T_F \gg \Theta_D$), and thus expected BCS superconducting properties, into question. This breakdown of the Frohlich approximation may explain the anomalous estimate of $\Delta =\hbar v_F/\pi \xi \sim$ 210 K, obtained from experimental parameters, $\xi=10$ nm and $v_F=8.6\times 10^5$ m/s.

Second, non-centrosymmetry of the half-Heusler crystal structure forces an automatic mixing of singlet and triplet pairing states in the superconductivity, resulting in not only complex superconducting gap functions, but also very unique excitations involved in the Majorana fermions in topological superconductors by controlling the singlet and triplet contributions. Indeed, in the non-centrosymmetric superconductor Li$_{2}$(Pt,Pd)$_{3}$B, by changing the relative concentrations of Pt and Pd, one can tune the spin parity of the gap function from singlet domiant to triplet dominant \cite{harad12}. 

Third, the coexistence of magnetism and superconductivity in this system not only adds another of only a few canonical magnetic superconductor platforms such as the borocarbide \cite{mulle01} and Chevrel phase \cite{fisch78}, but may also serve as a unique platform to investigate topological orders with multi-symmetry breaking states. 

Finally, we note that the tunability of the band structure via chemical means provides an ideal platform for traversing the quantum phase transition between trivial and non-trivial topological states, traversing the critical lattice parameter and tuning the interplay between antiferromagnetic and superconducting ground states continuously and controllably.
Our transport measurements are quite consistent with band calculations, which reveal that {\rpdbi} is a promising candidate for a topological semimetal. Photoemission studies of the related compound $R$PtBi ($R$ = Lu, Dy, and Gd) have observed metallic surface states not inconsistent with TI properties predicted by band calculations \cite{liu11}.

In summary, we have investigated the coexistence of magnetism and superconductivity in the single-crystal {\rpdbi} series. The magnetic rare earth members of this series exhibit an antiferromagnetic ordered state with $Q=(\frac{1}{2},\frac{1}{2},\frac{1}{2})$ due to RKKY interaction between conduction electrons and localized moments. Except for GdPdBi, all {\rpdbi} members exhibit bulk superconductivity with an unusual upper critical field behavior that suggests odd parity superconductivity, likely due to the lack of inversion symmetry in the crystal structure. The scaling of both magnetic order and superconductivity with de Gennes factor indicates {\rpdbi} to be a new family of magnetic superconductors, which together with the proposed topological properties provide a unique platform to investigate the emergence of novel quantum states of matter. %A unique coexistence of non-trivial topologies in both normal and superconducting state wavefunctions is possible, raising much theoretical curiosity.

%******************************
%METHODS
\section{Methods}

Single crystals of {\rpdbi} were grown using a Bi self-flux technique in a ratio of $R$:Pd:Bi=1:1:5-10. The typical size of resulting crystals is $\sim$3$\times$3$\times$3 mm$^3$. We confirmed the {\rpdbi} phase with powder x-ray diffraction, without any measurable impurity phase except very small traces of Bi flux (see Supplementary Material for more information). Low-temperature resistivity and AC susceptibility measurements were performed in a dilution refrigerator. The AC susceptibility signal was obtained in a driving field $H_{ac} = 0.1$ Oe. We measured high-temperature transport and heat capacity using the $^3$He option of a PPMS. DC magnetization was obtained using a MPMS. In the transport and magnetization measurements, we applied field along the [100] orientation. The neutron measurements were carried out on the BT-7 spectrometer \cite{lynn12} using collimations of 80$^\prime$ full-width-at-half-maximum before and after the pyrolytic graphite (PG) monochromator and 80$^\prime$ radial collimator after the sample, with the 1-d position sensitive detector. An incident monochromatic beam of $\lambda$ = 2.359 {\AA} was employed with a PG filter to suppress higher-order wavelengths. A pumped helium cryostat was used to control the sample temperature.

%*******************************************************

\section{acknowledgments}
We acknowledge useful discussions with K. Behnia, P.C. Canfield, J.C. Davis and M.B. Maple. Materials synthesis efforts were supported by DOE-Early Career award DE-SC-0010605, and characterization work was supported by AFOSR grant No. FA9550-14-1-0332. Competing Interests: The authors declare that they have no competing interests.

%****** FIGURES

\begin{figure*}[pb]
\includegraphics[width=16cm]{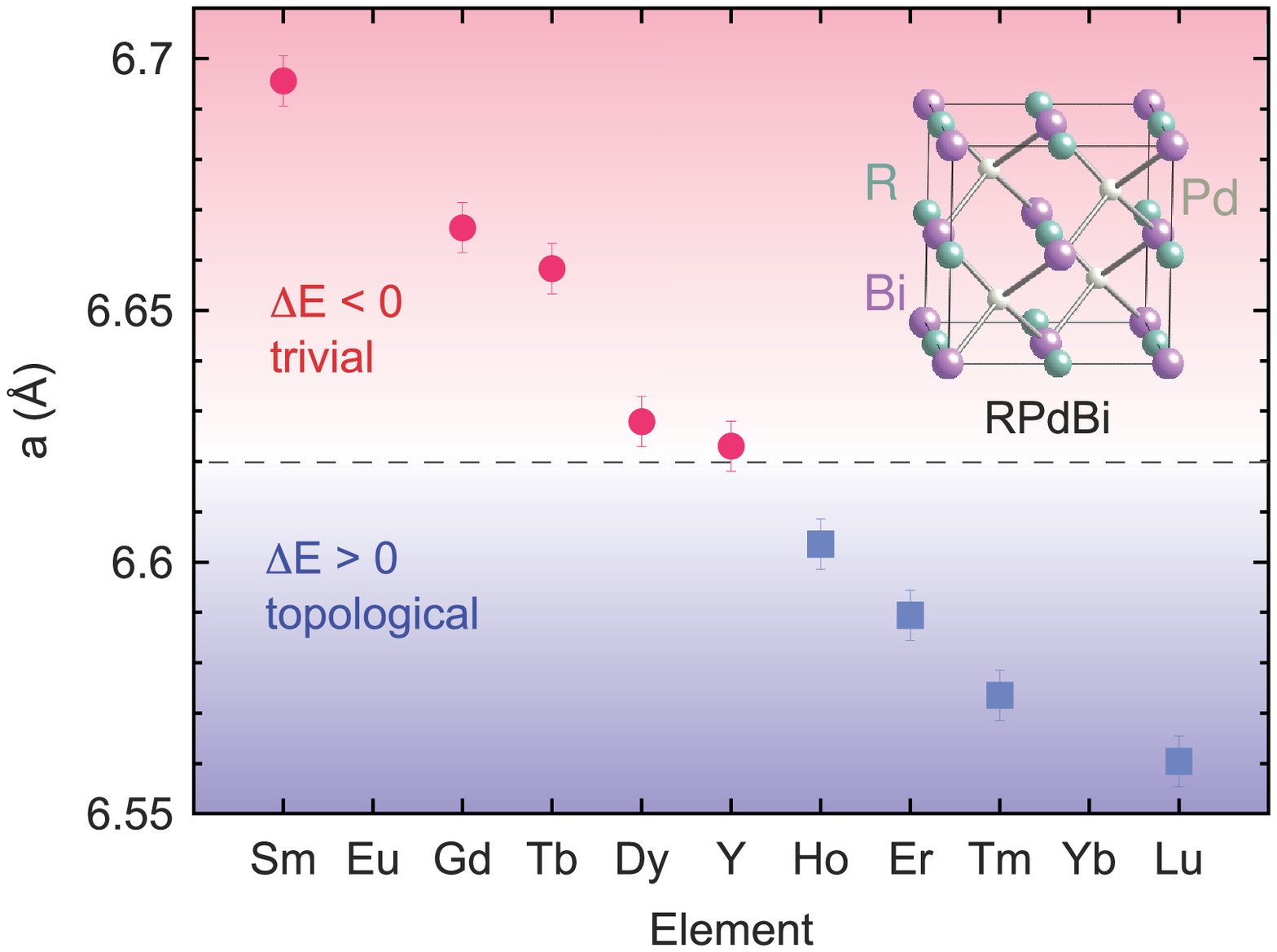}
\caption{Evolution of lattice constants as a function of rare earth species $R$ in {\rpdbi} determined by X-ray diffraction, showing the lanthanide contraction effect on the half-Heusler cubic ($F\bar{4}3m$) crystal structure (inset). The dashed line indicates the critical lattice constant $a_c = 6.62$ \AA\ demarcation between positive and negative electronic band inversion strength $\Delta E =E_{\Gamma 8} - E_{\Gamma 6}$ \cite{chado10}, and therefore between {\rpdbi} members with predicted topologically non-trivial (squares) and trivial (circles) band structure.}
\end{figure*}

\begin{figure*}[ptb]
\includegraphics[width=16cm]{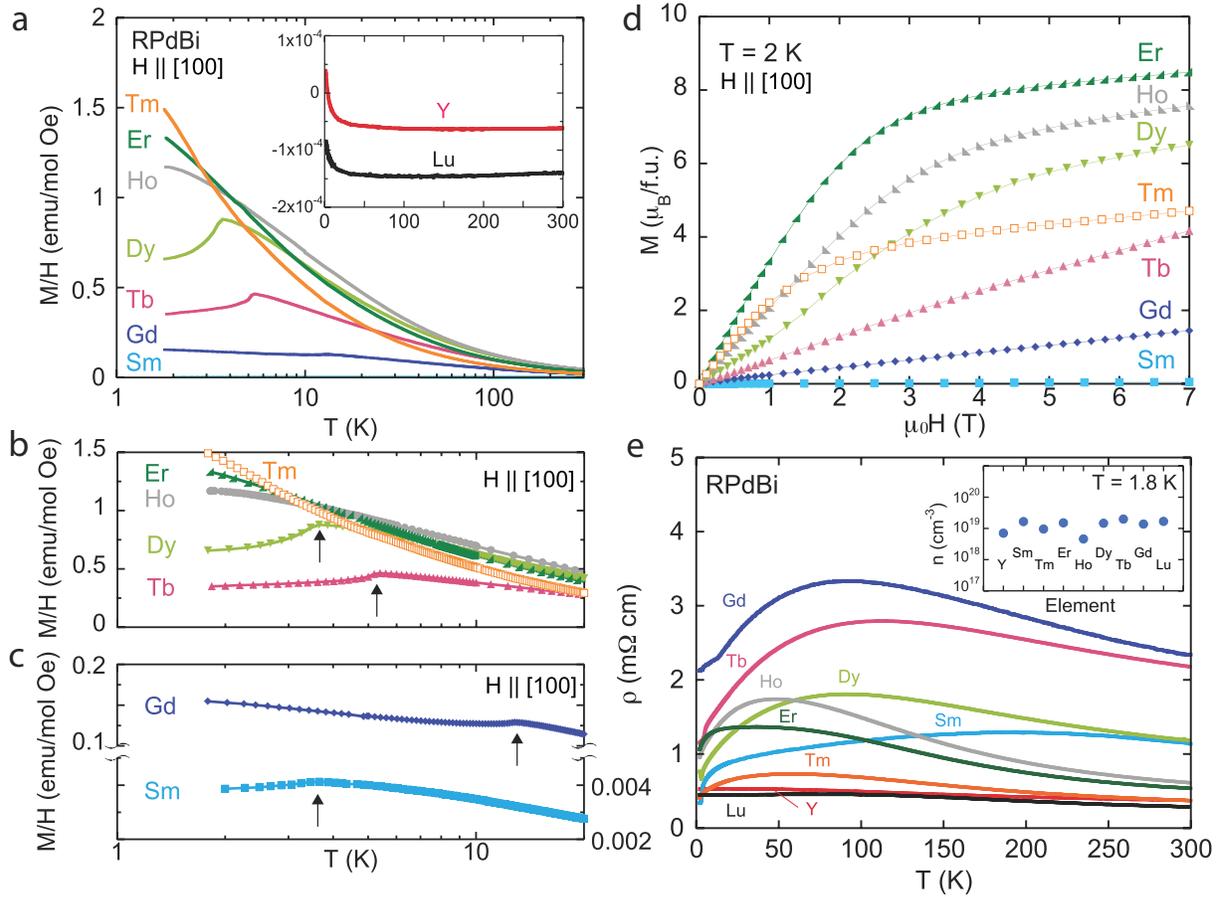}
\caption{Physical properties of {\rpdbi} single crystals, focusing on evolution of magnetic order with rare earth species $R$ = Sm, Gd, Tb, Dy, Ho, Er, and Tm. 
{\bf a} Magnetic susceptibility $M/H$ of {\rpdbi} members with magnetic $R$ species, showing Curie-Weiss behavior and clear, abrupt decreases in $M/H$ denoting antiferromagnetic transitions. Inset presents data for non-magnetic YPdBi and LuPdBi, exhibiting diamagnetic behavior. {\bf b} Low-temperature zoom of $M/H$ for Tb, Dy, Ho, Er, and Tm and {\bf c} for Gd and Sm, with arrows indicating N\'eel temperatures. (Note: 1~emu/mol~Oe=4$\pi\times$10$^{-6}$ m$^{3}$/mol.) {\bf d} Magnetization $M$ at 2~K for magnetic rare earth members $R$ = Sm, Gd, Tb, Dy, Ho, Er, and Tm. {\bf e}, Electrical resistivity of all members in the temperature range 2~K -- 300~K, showing non-monotonic temperature dependence in all species. The inset presents the charge carrier density $n_H$ obtained from single-band analysis of Hall effect measurements performed at 1.8 K (see text). The sign of all carriers is positive.}
\end{figure*}

\begin{figure*}[ptb]
\includegraphics[width=16cm]{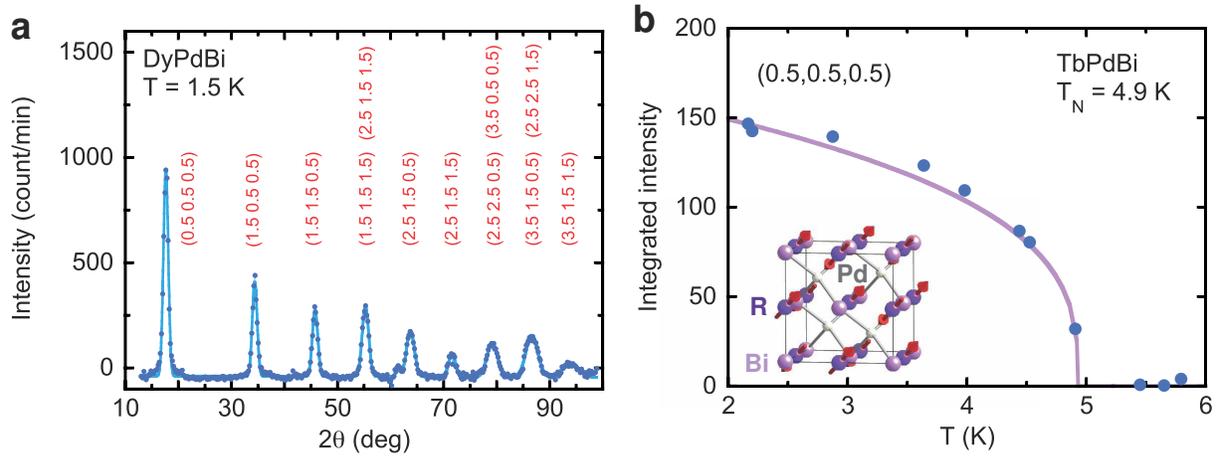}
\caption{Characterization of antiferromagnetic order with elastic neutron diffraction. 
{\bf a}, Low temperature magnetic diffraction pattern of DyPdBi obtained by subtracting 18~K data from 1.5~K data. Labels indicate the series of half-integer antiferromagnetic peaks. {\bf b}, antiferromagnetic order parameter of single-crystal TbPdBi obtained from the intensity of the (0.5,0.5,0.5) magnetic Bragg peak. Solid curve is a mean-field fit to the data, and the inset presents a schematic of the antiferromagnetic spin structure.}
\end{figure*}

\begin{figure*}[ptb]
\includegraphics[width=16cm]{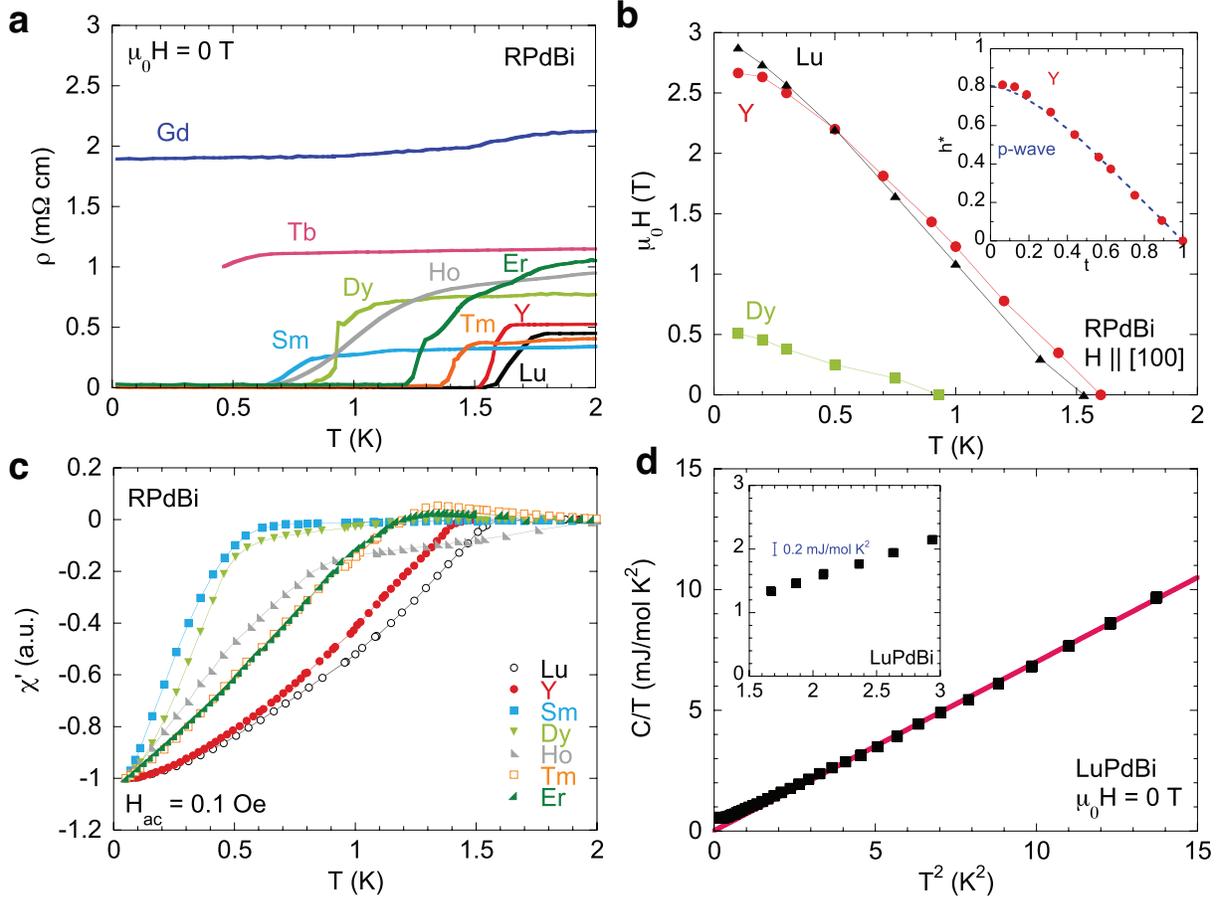}
\caption{Superconducting state properties of {\rpdbi} single crystals.
{\bf a}, Resistivity of {\rpdbi} at temperatures below 2~K, showing superconducting transitions for each compound. 
{\bf b}, Temperature dependence of upper critical field {\hc2} obtained from the resistive transition for YPdBi, LuPdBi and DyPdBi. The inset shows the normalized upper critical field $h^\ast=H_{c2}/T_c dH_{c2}/dT|_{T=T_c}$ as a function of normalized temperature $t=T/T_c$ for YPdBi, with dashed line indicating the expectation for a polar $p$-wave state. 
{\bf c}, AC susceptibility of single-crystal samples of {\rpdbi}. The magnitude of the screening below the superconducting transition is comparable to a test sample of superconducting aluminum, confirming bulk diamagnetic screening (see text). 
{\bf d}, heat capacity $C/T$ as a function of $T^{2}$ for LuPdBi. Solid line is a fit to the data using $C/T = \gamma + \beta T^{2}$, where $\gamma T$ is the electronic and $\beta T^3$ is the phonon contribution to the specific heat. Inset: enlarged view of the $C/T$ vs $T^2$ near $T_c\sim 1.5$ K, with estimated size of jump at the superconducting transition based on BCS expectation shown as an error bar (see text).}
\end{figure*}

\begin{figure*}[ptb]
\includegraphics[width=16cm]{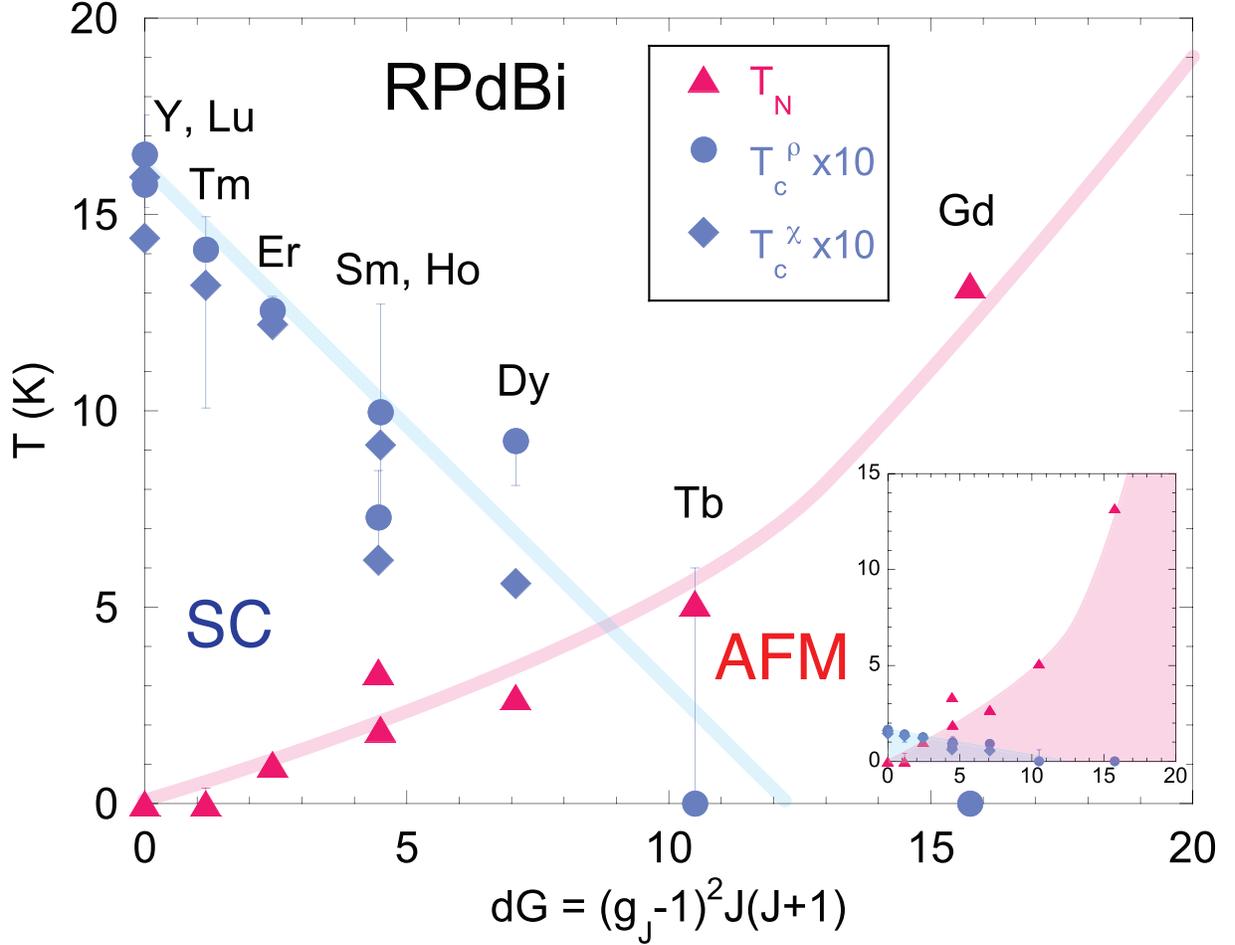}
\caption{Phase diagram of {\rpdbi} series, indicating evolution of superconducting (SC) and antiferromagnetic (AFM) ground states as a function of de Gennes factor $dG = (g_{J}-1)^{2}J(J+1)$. The superconducting transition {\tc} (blue) is obtained from 
the midpoint of the resistive transition (circles; upper and lower error bars indicate onset and zero resistance) and the onset of diamagnetism in AC susceptibility (diamonds), and N\'eel temperature {\tn} (red triangles) is obtained from DC magnetic susceptibility. The plotted {\tc} is scaled by a factor of 10, and solid lines are guides to the eye. Note that {\tc} ({\tn}) for SmPdBi is lower (higher) than that for HoPdBi. Inset: unscaled {\tc} and {\tn} as a function of $dG$.}
\end{figure*}

\bibliographystyle{apsrev}

\clearpage
\widetext

\begin{center}
\textbf{\large Supplementary materials for Topological {\rpdbi} half-Heusler semimetals: \\
a new family of non-centrosymmetric magnetic superconductors}
\end{center}

\newcommand{\beginsupplement}{%
        \setcounter{figure}{0}
        \renewcommand{\thefigure}{S\arabic{figure}}%
     }
\beginsupplement
\twocolumngrid

\section{Structural characterization}
The powder X ray diffraction patterns confirm the formation of {\rpdbi} phase with a MgAgAs-type structure (space group $F\bar{4}3m$) as shown in Fig.~S1. Except for Bi flux residues, the observation of single phase of {\rpdbi} ensures the observed magnetism and superconductivity are intrinsic. The obtained lattice parameters are close to the reported values \cite{haase02} (Fig.~1 in the main text).

\section{Exclusion of possible impurity phases}
Although X ray diffraction patterns show a single phase with small flux residues in our samples, it is worth excluding the possible contamination of extrinsic superconducting phases from the observed superconductivity. The possible extrinsic phases can be well-known binary superconductors consist of Pd and Bi. The binary alloys have different crystal structures with different $T_c$, namely, monoclinic $\alpha$-BiPd (space group $P2_1$) with $T_c =$ 3.8 K and $\mu_0 H_{c2}(0) = 0.7$ T \cite{joshi11}, monoclinic $\alpha$-Bi$_2$Pd ($C2/m$) with $T_c =$ 1.7 K \cite{matth63a}, and tetragonal $\beta$-Bi$_2$Pd ($I4/mmm$) with $T_c =$ 5.4 K and $\mu_0 H_{c2}^{ab}(0)~(\mu_0 H_{c2}^{c}(0)) = 1.1~(0.7)$ T \cite{imai12}. Since the observed transition temperature in {\rpdbi} does not exceed 1.6 K as shown in Fig.~5 in the main text, we can easily exclude the contamination of monoclinic $\alpha$-PdBi and tetragonal $\beta$-PdBi$_2$ with much higher $T_c$. On the other hand, $T_c$ of $\alpha$-Bi$_2$Pd, rather close to the observed $T_c$ of {\rpdbi}, requires further investigation as described the following paragraph. 

To have a closer look at the superconducting properties of $\alpha$-Bi$_2$Pd, we present the resistivity of single crystals obtained from flux method in Fig.~S2a. Resistivity shows metallic behavior on cooling, followed by the superconducting transition at 1.3 K, slightly lower than the reported $T_c$ \cite{matth63a}. The superconductivity is strongly suppressed by magnetic field, and the critical field obtained from the resistive transition is approximately 300 Oe along out-of- and in-plane orientations at 400 mK (Fig.~S2b). The extremely small critical field in monoclinic $\alpha$-Bi$_2$Pd is strongly suggestive of inconsistency with the observed $\mu_0 H_{c2}$ of $\sim$3 T for YPdBi and LuPdBi and $\sim$0.5 T for DyPdBi. 

The discrepancy of the transition temperature and critical field indicates the observed superconductivity in {\rpdbi} is intrinsic, not due to the contamination of Pd-Bi superconducting alloys. Besides, we note that amorphous Bi undergoes superconductivity at $T_c=6$ K \cite{matth63a}, also inconsistent with the observed $T_c$ in {\rpdbi}. Interestingly, heat treatments at $\sim 200^{\circ}$C induce surface superconductivity with $T_c$ of 1.6 K in all {\rpdbi}, independent of the de Gennes factor, which cannot be explained by the formation of a secondary phase and may be correlated with the topological nature. 

\section{heat capacity}
Besides the resistivity and magnetic susceptibility shown in the Fig.~2, low temperature heat capacity also reveals the magnetic transition for {\rpdbi} except for Tm and non-magnetic Y and Lu as shown in Fig.~S3. The heat capacity jump at N\'eel temperature $T_N$ is gigantic, and its magnitude is the order of J/mol K$^2$, suggesting a huge release of magnetic entropy. Note that upon cooling to 400 mK, the heat capacity for TmPdBi continues to increase, indicating a possible precursor of a magnetic transition below the base temperature of 400 mK. We list the obtained $T_N$ in TABLE S1.

\section{Magnetic properties}
As shown in Fig.~2 in the main text, the temperature dependence of the susceptibility for magnetic rare earth {\rpdbi} shows Curie-Weiss behavior at high temperatures, indicating that the magnetic properties originate from localized $4f$ electron moments. From fitting of the Currie-Weiss law, we obtained the effective moments of $4f$ electrons, $\mu_\mathit{eff}$, close to the theoretical ones for respective free $R^{3+}$ ions, $\mu_\mathit{free}$, summarize in TABLE S1. The obtained Weiss temperatures $\Theta_{W}$ are negative, consistent with the observed antiferromagnetic orderings in this system.

\begin{table*}[p]
\begin{tabular}{c|cccc}
\hline\hline
$R$ & $T_N$ (K) & $\Theta_{W}$ (K) & $\mu_\mathit{eff}$ ($\mu_B$) & $\mu_\mathit{free}$ ($\mu_B$)\\
\hline
Sm & 3.4 & -258 & 1.9 & 0.85 \\
Gd & 13.2 & -49.6 & 7.66 & 7.94 \\
Tb & 5.1 & -28.9 & 9.79 & 9.72 \\
Dy & 2.7 & -14.3 & 10.58 & 10.65 \\
Ho & 1.9 & -9.4 & 10.6 & 10.6 \\
Er & 1.0 & -4.8 & 9.18 & 9.58 \\
Tm & $<$0.4 & -1.7 & 7.32 & 7.56 \\
\hline\hline
\end{tabular}
\caption{N\'eel temperature $T_N$ obtained from the heat capacity and magnetization, Weiss temperature $\Theta_{W}$, and the effective moments $\mu_\mathit{eff}$ for {\rpdbi} obtained from a fit to the Curie-Weiss expression. The obtained effective moments are close to the free ion moments $\mu_\mathit{free}$ except for Sm.}
\end{table*}

\begin{figure*}[p]
\includegraphics[width=17cm]{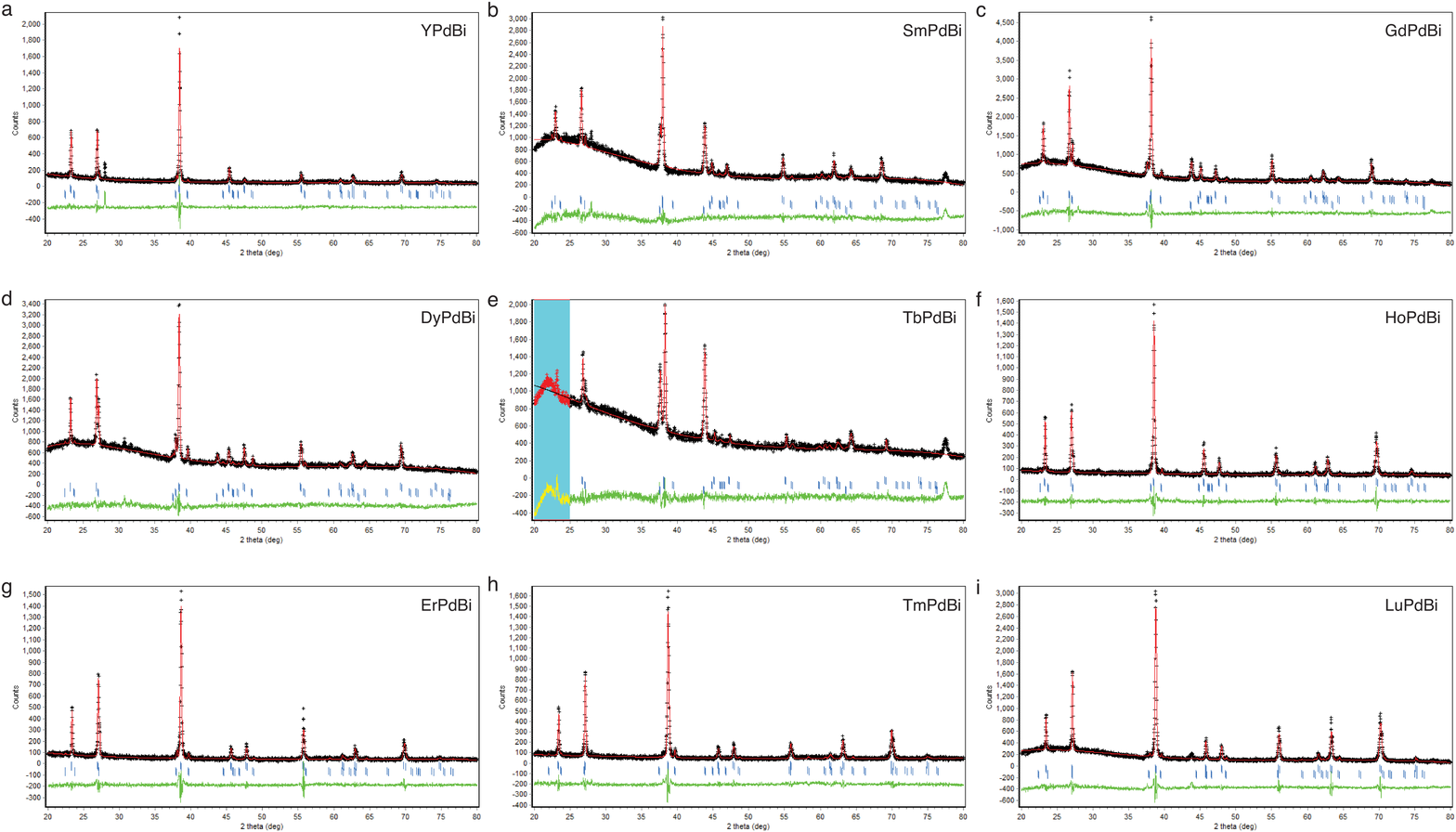}%
\caption{{\bf X ray diffraction patterns for {\rpdbi} with Cu $K\alpha$ radiation.} Cross symbols are observed data, and lines are theoretical calculations. Upper markers and lower makers correspond to reflections for RPdBi and for elemental Bi. For Sm, Gd, Dy, and Tb, we used two different Bi with $R\bar{3}m$ (middle markers) and $Fm\bar{3}m$ (bottom markers) for the refinements.}
\end{figure*}

\begin{figure*}[p]
\includegraphics[width=12cm]{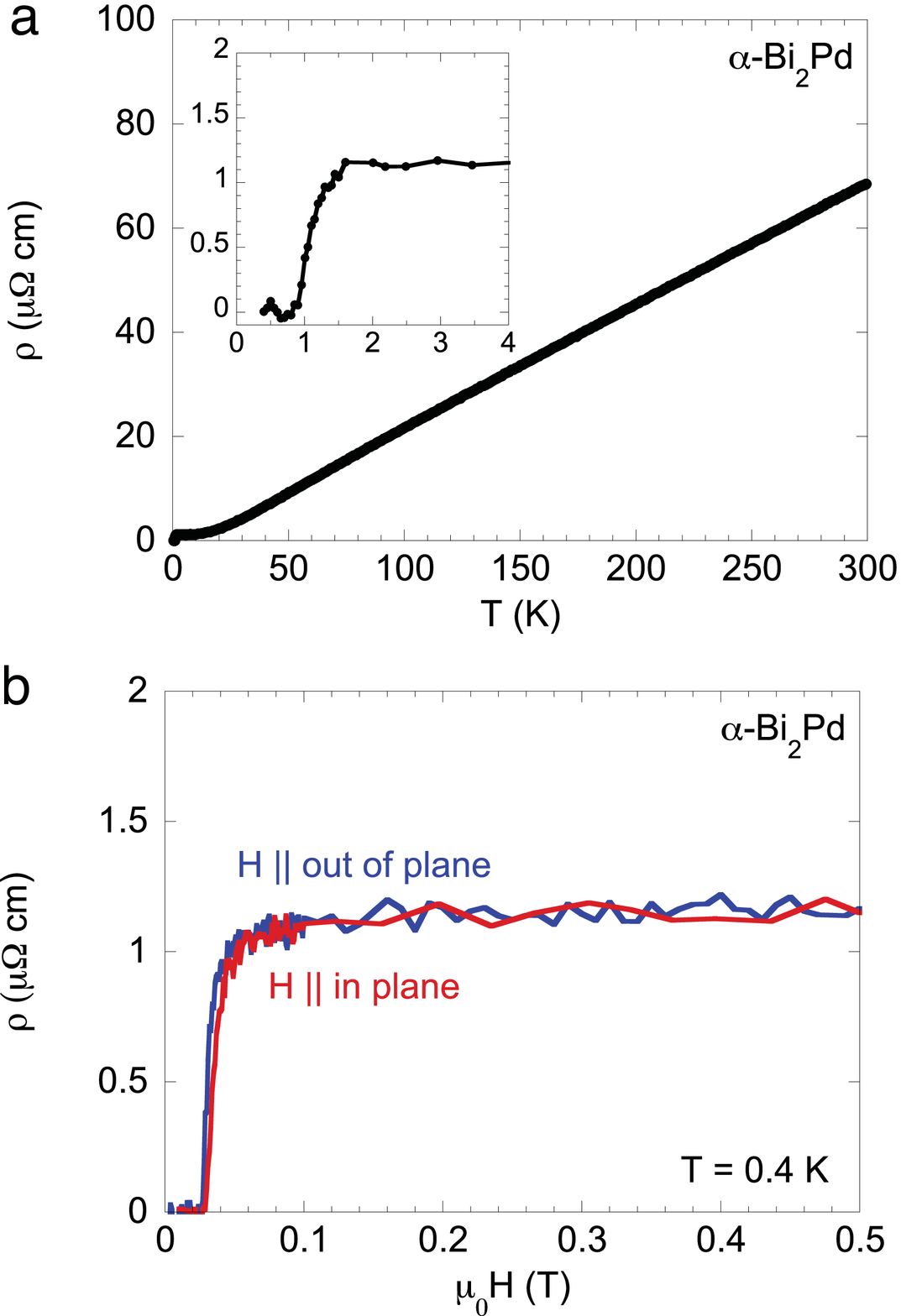}
\caption{{\bf Transport properties of {\pdbi2}.} {\bf a}, Resistivity as a function of temperature for {\pdbi2}. Inset: low temperature part of the resistivity. {\bf b}, Field dependence of resistivity.}
\end{figure*}

\begin{figure*}[p]
\includegraphics[width=17cm]{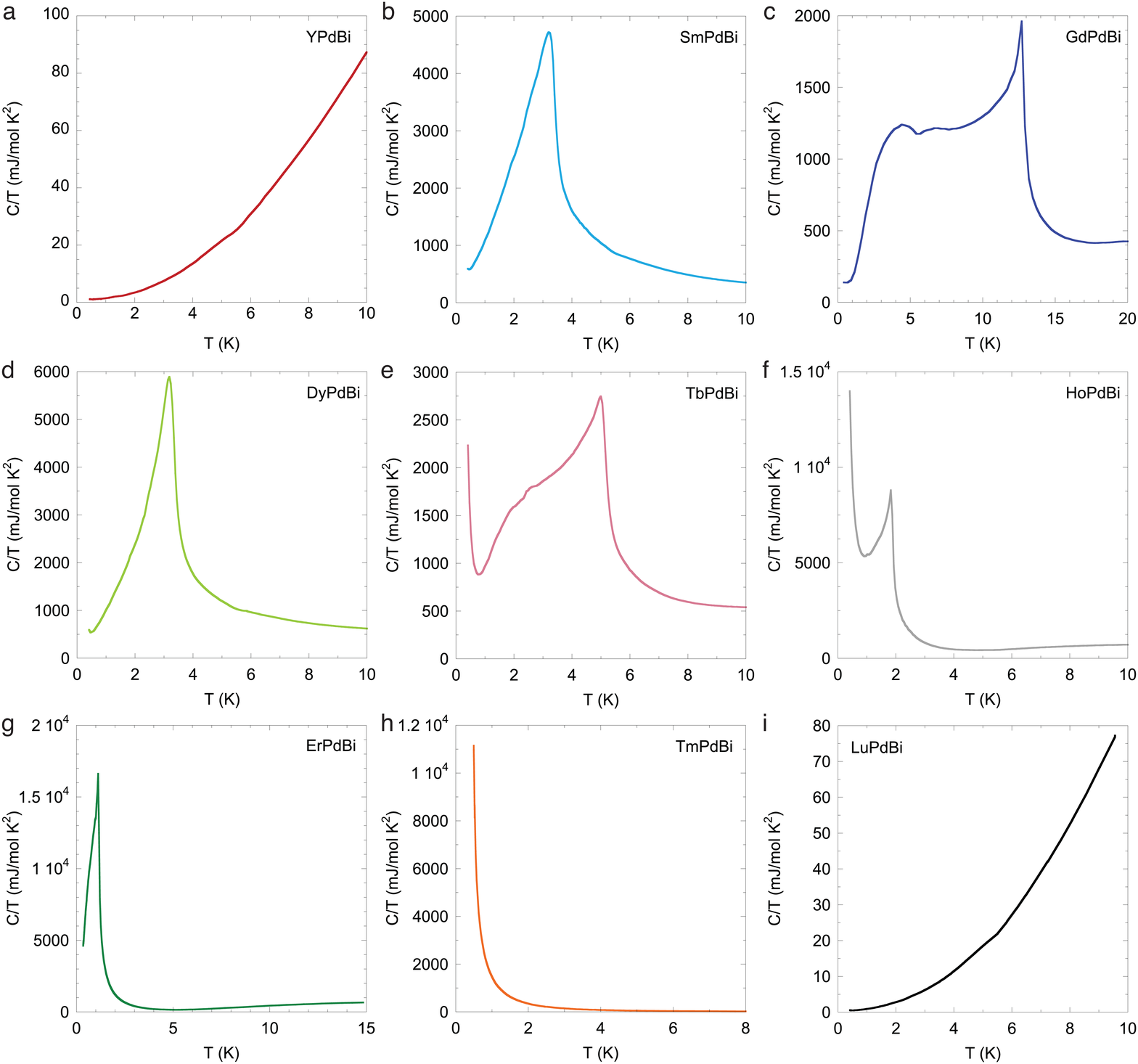}%
\caption{{\bf Specific heat as a function of temperature for {\rpdbi}.} Except for non-magnetic Y and Lu, {\rpdbi} undergo antiferromagnetic order at low temperatures. The steep increase in the heat capacity of TmPdBi at low temperatures suggests a possible precursor of magnetic order below 0.4 K.}
\end{figure*}

\end{document}